\documentclass{agujournal}

\journalname{JGR-Space Physics}

\begin{document}

\title{Electron Physics in 3D Two-Fluid Ten-Moment Modeling of Ganymede's
Magnetosphere}

\authors{Liang Wang\affil{1}}
\authors{Kai Germaschewski\affil{1}}
\authors{Ammar Hakim\affil{2}}
\authors{Chuanfei Dong\affil{2,3}}
\authors{Joachim Raeder\affil{1}}
\authors{Amitava Bhattacharjee\affil{2,3}}

\affiliation{1}{Space Science Center, University of New Hampshire, Durham, NH
03824, USA}
\affiliation{2}{Princeton Center for Heliophysics, Princeton Plasma Physics
Laboratory, Princeton University, Princeton, NJ 08544, USA}
\affiliation{3}{Department of Astrophysical Sciences, Princeton University,
Princeton, NJ 08544, USA}

\correspondingauthor{L. Wang}{liang.wang@unh.edu}

\begin{keypoints}
\item Non-gyrotropic electron pressure tensor effects are important in
Ganymede's magnetic reconnection
\item Electrons and ions form highly asymmetric and distinct drift patterns in
the magnetosphere
\item Some key features of the observed oxygen emission morphologies are
reproduced by the model
\end{keypoints}

\begin{abstract} We studied the role of electron physics in 3D two-fluid
10-moment simulation of the Ganymede's magnetosphere. The model captures
non-ideal physics like the Hall effect, the electron inertia, and anisotropic,
non-gyrotropic pressure effects. A series of analyses were carried out: 1) The
resulting magnetic field topology and electron and ion convection patterns
were investigated. The magnetic fields were shown to agree reasonably well
with in-situ measurements by the Galileo satellite. 2) The physics of
collisionless magnetic reconnection were carefully examined in terms of the
current sheet formation and decomposition of generalized Ohm's law. The
importance of pressure anisotropy and non-gyrotropy in supporting the
reconnection electric field is confirmed.
3) We compared surface ``brightness'' morphology, represented by surface
   electron and ion pressure contours, with oxygen emission observed by the
   Hubble Space Telescope (HST). The correlation between the observed emission
   morphology and spatial variability in electron/ion pressure was
   demonstrated.  Potential extension to multi-ion species in the context of
   Ganymede and other magnetospheric systems is also discussed.
\end{abstract}

\section{Introduction}

Ganymede, a moon of Jupiter, is not only the largest satellite in the solar
system, but also the only satellite that possesses an internal dynamo to
generate an intrinsic dipole magnetic field
\citep{Gurnett1996,Kivelson1996,Kivelson1997,Kivelson2002}. This dipole field
interacts with the corotating Jovian plasma and magnetic field, much like the
Earth's dipole interacts with the solar wind in the southward $B_{z}$
situation, to form Ganymede's own magnetosphere embedded in the enormous
magnetosphere of Jupiter. The ``incident'' Jovian plasma flow is
sub-magnetosonic, thus no bow shock is formed around the moon. Instead, a pair
of tube-like structures, called Alfv\'en wings, are induced at the moon,
within which the incident plasmas are greatly slowed down
\citep{Kivelson1997}.

Ganymede's magnetosphere is relatively small in terms of the moon's own size
and kinetic scales like particle inertia lengths and gyroradii. This makes it
an ideal numerical laboratory for high-resolution investigation of various
topics in magnetospheric physics. Previously, the Ganymede's magnetosphere has
been modeled using various models. Kopp and Ip performed single-fluid MHD
simulations to reveal the global structure of Ganymede's field line topology,
current system, and convection pattern under different upstream Jovian
conditions \citep{Kopp2002a,Ip2002}. A series of MHD studies by Jia et al.
focused on the details of magnetic reconnection like the formation of Flux
Transfer Event (FTE) and discussed its unsteady nature
\citep{Jia2008,Jia2009,Jia2010}. They also compared different inner boundary
conditions at the moon's surface, and were able to achieve good agreement with
in-situ plasma and field measurement when they treated the inner boundary as a
plasma sink with finite conductance. \citet{Duling2014} developed a set of rigorous
boundary conditions for their MHD simulations, including different conditions
for the poloidal and toroidal components of the magnetic field. Their simulations further uncovered the presence of a
salty conductive ocean beneath Ganymede's surface. Paty et al. used multi-ion
MHD model to study how different ion species are supplied at the ionosphere
and convected and energized in the magnetosphere
\citep{Paty2004,Paty2006,Paty2008}. They further developed a brightness model
to understand the morphology of the Ganymede's auroral oval
\citep{Payan2015a}. Later, \citet{Dorelli2015} showed that introduction of Hall
effect in an MHD model produced significant asymmetry in plasma flow paths and
modified the local field line topology. More recently,
\citet{Fatemi2016} used a hybrid model (ions as discrete particles and
electrons as massless fluid) to calculate Jovian ion fluxes precipitating on
the surface of Ganymede and their correlation to Ganymede's surface brightness
.

A less explored topic is the role of electron kinetic effects. Numerous
theoretical analyses, local fully kinetic Particle-In-Cell (PIC) simulations
and in-situ observations have confirmed the importance of electron kinetic
effects in collisionless magnetic reconnection
\citep{Vasyliunas1975,Cai1997,Kuznetsova1998,Kuznetsova2001a,Birn2001,Wang2000,ma1998grl,Oieroset2001}.
In highly collisionless space plasmas, collisional resistivity is negligible,
thus reconnection electric field at the X-line, an important indicator of how
fast reconnection is taking place, has to be supported by collisionless
effects like the divergence of non-gyrotropic electron pressure tensor
\citep{Vasyliunas1975,Cai1997,Kuznetsova1998,Kuznetsova2001a,Birn2001,Oieroset2001}.
Since magnetic reconnection is one of the major mechanisms of fast energy
conversion and release, its rate can significantly affect the overall
structure and evolution of the magnetosphere. How to accurately and
efficiently incorporate necessary electron kinetic effects has been a
long-standing issue in global modeling of various magnetospheric systems,
given the fact that full-domain fully kinetic simulations are much too
expensive with contemporary computing power.

One notable attempt to address this issue is the MHD-EPIC approach suggested
by \citet{Daldorff2014}, which solves MHD or Hall MHD equations in the
majority of the domain, but treats plasma as discrete particles (using the
implicit PIC method) in prescribed regions where kinetic effects are
potentially important. \replaced{This model has been successfully applied to
the Ganymede's magnetosphere, and demonstrated significant differences
compared with standalone Hall MHD results {[\textit{T\'oth et al.}, 2016].}}
{The MHD-EPIC model has been successfully applied to the Ganymede's
magnetosphere \citep{Toth2016}. The overall structure of the
magnetosphere was similar to that from
Hall MHD solution, indicating that kinetic effects do not
significantly change the global configuration. But the MHD-EPIC solutions were generally more dynamic than
the Hall MHD solutions, producing substantially more and larger FTEs. The
MHD-EPIC results also agreed better with Galileo observations in terms of time
series and fluctuation spectrum of magnetic field. Beyond the analyses
performed by \citet{Toth2016}, the MHD-EPIC model can also be used to study
particle acceleration, wave-particle interaction, etc., due to the fully
kinetic nature of its PIC component.}. However, \citet{Toth2016} focused more
on the general validation of the model and on FTE formation, and did not
include dedicated discussion on the roles of electron physics. Their PIC
region grid did not fully resolve electron inertia length
($\Delta_{\mathrm{minimum}}\approx0.8d_{e,\mathrm{Jovian}}$), either, thus the
electron effects might have not been accurately modeled. In addition, kinetic
dynamics are handled in a few localized box regions to minimize the cost of
fully kinetic calculation, but the magnetospheric convection and field-aligned
current system are often of global scale, which might also require kinetic
treatment.

In this paper, we suggest an alternative approach to incorporate electron (in
addition to ion) kinetic effects using higher-order moment multi-fluid models.
\replaced{This model evolves moment equations of each species in the plasma,
including both electrons and ions {[\textit{Hakim et al.}, 2006;
\textit{Hakim}, 2008; \textit{Wang et al.}, 2015]}. Particularly, we focus on
the 10-moment sub-model which evolves full pressure tensors, without the
isotropic or gyrotropic assumptions.}{ These models solve a hierarchy of
velocity-space moment equations truncated at a given order. Depending on the
number of moment equations retained, we obtain different
``sub-models''\citep{Hakim2006,Hakim2008,Wang2015a}. Particularly, we focus on
the 10-moment sub-model which evolves the full pressure tensor, in addition to
density and momentum, without any isotropy or gyrotropy assumptions.  Totally
ten independent moment terms (one density, three momenta, and six independent
pressure tensor elements) are evolved for each species in the 10-moment
sub-model, hence the name (see Section~\ref{sec:model} for more details).}
This way, the electron and ion densities, momenta, and pressure tensors are
both directly evolved in time to account for their separate dynamics.
\replaced{Another consequence is that the generalized Ohm’s law embedded in
the species momentum equations then automatically contains non-ideal effects
like the Hall effect, electron inertia, and divergence of electron pressure
tensor, to support the reconnection electric field.}{Another consequence is
that the same physics which are conventionally put into a generalized Ohm's
law are automatically fully included by means of solving all species' momentum
equation directly, coupled via Maxwell's equations. In particular, the model
contains a Hall term, electron pressure tensor and inertia effects.} Due to the
fluid nature of the model, its computational cost is more comparable to traditional
MHD-based codes.

We will perform two-fluid (electron-ion) 10-moment simulations to study the
dynamics of both electron and ion flows and the role of electron physics in
collisionless reconnection. Key features of the Ganymede's magnetosphere,
particularly the asymmetry in global flow
\replaced{patters}{patterns} and formation of the auroral oval, will also be
investigated. We will not focus on the ionospheric outflow, though, since a
relatively simple inner boundary condition is employed, similar to that of
\citep{Dorelli2015,Toth2016}. The consequences due to this limitation will
also be briefly discussed.

The paper is outlined as follows: The equations of the 10-moment model are
introduced in Section \ref{sec:model}. Initial setup, boundary conditions, and
various parameters for the Ganymede simulations are described in Section
\ref{sec:setup}. The simulation results are presented and discussed in Section
\ref{sec:Results}, followed by a summary of findings and implications for
future work in Section~\ref{sec:Conclusions}.

\section{\label{sec:model}The 10-Moment Model Equations}

The high-moment multi-fluid model is constructed by taking velocity moments of
Vlasov equations of each species, including electron, to obtain a hierarchical
set of moment equations. Closure is required at the highest truncated
moment. For example, in the 10-moment sub-model that truncates at the 2nd
order moment, the pressure tensor, assumptions should be made to approximate
the 3rd order moment, the heat-flux tensor, depending on the problems studied.
In this manner, the high-moment multi-fluid model can, in principle,
incorporate even higher-order of moments and thus give a more complete kinetic
representation of the plasma than the traditional two-fluid or single-fluid
(MHD) model with scalar pressure. Different from MHD-based models which assume
charge neutrality and derive electron dynamics indirectly, the high-moment
multi-fluid model treats electrons as an independent species and track its
evolution. In addition, support of multiple ion species becomes
straightforward since we simply need to integrate more sets of moment
equations using the same numerical solver.

In this work, we will use the two-fluid (one electron species and one ion
species) 10-moment model to model the Ganymede's magnetosphere. A total of ten
equations are solved for each species:
\begin{eqnarray}
\frac{\partial\left(m_{s}n_{s}\right)}{\partial
t}+\frac{\partial\left(m_{s}n_{s}u_{j,s}\right)}{\partial x_{j}} & = & 0,\\
\frac{\partial\left(m_{s}n_{s}u_{j,s}\right)}{\partial
t}+\frac{\partial\mathcal{P}_{ij,s}}{\partial x_{j}} & = &
n_{s}q_{s}\left(E_{i}+\epsilon_{ijk}u_{j,s}B_{k}\right),\label{eq:10m-momentum}\\
\frac{\partial\mathcal{P}_{ij,s}}{\partial
t}+\frac{\partial\mathcal{Q}_{ijm,s}}{\partial x_{m}} & = &
n_{s}q_{s}u_{[i,s}E_{j]}+\frac{q_{s}}{m_{s}}\epsilon_{[iml}\mathcal{P}_{mj],s}B_{l}.\label{eq:10m-pressure}
\end{eqnarray}

Here, subscripts $s=e,i$ represent the electron and ion species. They will be
neglected hereinafter for convenience. The square brackets in equation
(\ref{eq:10m-pressure}) around indices $ijm$ represent the minimal sum over
permutations of these indices that give completely symmetric tensors. The
first, second, and third order moments are defined as $mu_{i}\equiv m\int
v_{i}fd\mathbf{v}$, $\mathcal{P}_{ij}\equiv m\int v_{i}v_{j}fd\mathbf{v}$, and
$\mathcal{Q}_{ijm}\equiv m\int v_{i}v_{j}v_{m}fd\mathbf{v}$, with $f$ being
the phase space distribution function. For completeness, $\mathcal{P}_{ij}$
relates to the commonly used thermal pressure tensor
$P_{ij}\equiv\int\left(v_{i}-u_{i}\right)\left(v_{j}-u_{j}\right)fd\mathbf{v}$
by
\[
\mathcal{P}_{ij}=P_{ij}+nmu_{i}u_{j}.
\] $\mathcal{Q}_{ijm}$ relates to the conventionally defined heat flux tensor
$Q_{ijm}\equiv\int(v_{i}-u_{i})(v_{j}-u_{j})(v_{m}-u_{m})fd\mathbf{v}$
by
\[
\mathcal{Q}_{ijm}=Q_{ijm}+u_{[i}\mathcal{P}_{jm]}-2nmu_{i}u_{j}u_{m}.
\]

Once closed by an approximation to $Q_{ijk}$, the 10-moment equation system
self-consistently evolves the pressure tensor for each species. In this paper,
we adopt the following simple 3D closure devised by \citet{Wang2015a}:
\begin{equation}
\partial_{m}Q_{ijm}\approx
v_{t}\left|k\right|\left(P_{ij}-p\delta_{ij}\right).\label{eq:10-q-closure}
\end{equation} Here, the scalar wave number $k$ is one over a characteristic
length scale over which the dominant local physics takes place, $v_{t}$ is the
local thermal speed and $p\equiv\left(P_{xx}+P_{yy}+P_{zz}\right)/3$ is the
local scalar pressure. In 2D simulation of anti-parallel reconnection
presented by \citet{Wang2015a}, $k$ was chosen to be a constant, $1/d_{e0}$,
by arguing that the dominant physics, magnetic reconnection, takes place on
the length scale of upstream electron inertia length, $d_{e0}$. In this paper,
we relax this requirement by re-calculating $k_{s}$ as $1/d_{s}$ every time
step, where $d_{s}$ is the local inertia length of species $s$ at the time
step. This should provide more accurate heat flux approximation as species
inertia length in the Ganymede problem can vary greatly in space. More
recently, \citet{Ng2017} implemented equation (\ref{eq:10-q-closure}) in the
Fourier transform space ($k-$space) so that the solution in real space
($x$-space) can be recovered by performing \replaced{invert}{an inverse}
Fourier transform. This Fourier transform approach captures
non-local heat flux contribution and gives even better agreement with fully
kinetic simulation results. However, this approach requires Fast Fourier
transform (FFT), which is usually computationally too expensive for large
scale systems, and thus is not employed in our magnetosphere modeling here.

The electromagnetic field is evolved by full Maxwell equations
\begin{eqnarray}
\frac{1}{c^{2}}\frac{\partial\mathbf{E}}{\partial t} & = &
\nabla\times\mathbf{B}-\mu_{0}\mathbf{J},\label{eq:maxwell-dE}\\
\frac{\partial\mathbf{B}}{\partial t} & = & -\nabla\times\mathbf{E},
\end{eqnarray} thus electromagnetic waves are fully supported, similar to the
PIC and Vlasov models. An important difference from traditional MHD-based
models is that, the electric field $\mathbf{E}$ in high-moment multi-fluid
model is no longer computed explicitly from a given form of the Ohm's law
(e.g., \replaced{equation (2) of {[\textit{Dong et al.}, 2014]}}
{\citep{Vasyliunas1975,Birn2001}}), but is evolved in time by the Ampere's law
(\ref{eq:maxwell-dE}). The generalized Ohm's law \replaced{is embedded in}{can
be derived from} the equation system to include all orders of kinetic effects
retained in the moment equations and can be conceptually recovered by
combining the electron and ion momentum equations (\ref{eq:10m-momentum}).

By including the electron inertia, the dispersive modes (e.g., the whistler)
obey the exact dispersion relation which does not produce the quadratic
divergence of the oft-quoted dispersion relation obtained by excluding the
electron inertia. Thus the explicit time step is not restricted as the
dispersive modes grow, unlike inertia-less Hall MHD. The restriction due to
Debye length and plasma frequency can also be avoided by using a locally
implicit algorithm described by \citet{Hakimgkeyll}. Since the restriction due
to sound speed is usually relatively mild, the time step is often restricted
by speed of light only. For our studies, this restriction can also be
efficiently relaxed by using an artificially reduced speed of light.

The high-moment multi-fluid model has been successfully implemented in the
computational plasma physics framework, \emph{Gkeyll}. Non-uniform
``stretched'' Cartesian grid is supported. The high-resolution wave
propagation scheme, a variant of the finite-volume method, is used to solve
the hyperbolic equations. The code has been used for many laboratory and space
plasma physics projects. For details on the numerics and benchmark examples,
please refer to \citep{Hakim2006,Hakim2008,Wang2000,Hakimgkeyll}.

\section{\label{sec:setup}Numerical Setup}

We adopt a Cartesian coordinate system following the Ganymede-centered
Cartesian system (GphiO) definition \citep{Kivelson1997}: $+x$ is the inflow
direction of the corotating Jupiter plasma, $+y$ points from Ganymede to
Jupiter, and $+z$ is along the rotation axis. Ganymede is at the origin, and
the domain extends from $-64R_{G}$ to $64R_{G}$ in each direction, where
$R_{G}=2634.1{\rm km}$ is the radius of Ganymede. The domain is sufficiently
large to avoid unphysical reflection to significantly affect physics near the
Ganymede during the simulation. A stretched Cartesian $592\times576\times576$
grid is employed, with high resolution $\Delta_{\mathrm{minimum}}\approx
R_{G}/51$ in the box $\left[-2.5R_{G},2.5R_{G}\right]^{3}$. The grid size is
then smoothly ramped up to $1R_{G}$ at $\left|x,y,z\right|=15R_{G}$ and is
kept constant further out. We follow the convention in \citep{McGrath2013} to
call the $x<0$ side the upstream or (orbital) trailing side, the $x>0$ side
the downstream or (orbital) leading side. They are similar to the definitions
of day and night sides at the Earth's magnetosphere.

The initial magnetic field configuration is defined by the so-called ``mirror
dipole'' setup \added{(for details, see section 2.3 of \citep{Raeder2003} or
Text S1 in the Supporting Information}. In order to avoid large numerical
truncation errors, the strong intrinsic dipole field, $\mathbf{B}_{0}$, is
treated as a stationary background and does not contribute to the
$\nabla\times\mathbf{B}$ term in the Ampere's law. Only the fluctuating part
of the magnetic field, $\mathbf{B}_{1}$, is evolved and included in the
$\nabla\times\mathbf{B}$ term. Such separation of background dipole field and
induced perturbation field has been adopted in multiple global simulation
codes to improve numerical stability \citep{Tanaka1994,Toth2008,Janhunen2012}.
Initially, the plasma density and pressure are uniform throughout the domain
with inflow Jovian values, but the velocity is ramped from inflow velocity
down to zero over a spherical shell, $2R_{G}<r<2.5R_{G}$. The inflow Jovian
wind parameters, intrinsic dipole strength, and other parameters are
summarized in Table \ref{tab:ganymede-params}. According to these parameters,
the high resolution region resolves the inflow inertia length,
$d_{i,in}=\sqrt{m_{i}/\mu_{0}n_{i,in}\left|e\right|^{2}}\approx474km\approx0.18R_{G}$,
by about $9.2$ cells, and \added{marginally} resolves electron inertia length, $d_{e,in}$, by
about $1.83$ cells.

The inflow boundary at $x=-64R_{G}$ is set to fixed Jovian values listed in
Table \ref{tab:ganymede-params}. All other outer boundaries are set to float
(zero-gradient outflow). The inner boundary conditions at surface of Ganymede
are as follows: 1) plasma mass densities and pressures are set to constant
inflow values; 2) plasma momenta are radially reflected to achieve zero
in/out-flow in the steady state;
3) electric field is set to zero; 4) the fluctuating part of the magnetic
   field $\mathbf{B}_{1}$ is floated. Note that the inner boundary conditions
   we adopted are simpler than those used by \citep{Jia2009} and might not
   achieve as accurate agreement with details of in-situ measurements.
   Nevertheless, as we will show, the relatively simple boundary conditions
   are sufficient to correctly reproduce the overall structure of the
   magnetosphere that agrees reasonably well with observations.

\added{The simulation was performed on Trillian, a Cray XE6m-200 supercomputer
at UNH. A total of about 200,000 core hours were used to simulate about
15 minutes of physical time. This is much higher than the cost of a typical
single-fluid MHD simulation using the same grid size and time step size, due
to its larger number of equations to solve, but would still be significantly
cheaper than a fully kinetic simulation of a similar setup.}

\begin{table*}
\caption{\label{tab:ganymede-params}Parameters used in the Ganymede simulation
(adapted from \citep{Kivelson1997,Jia2010}). Terms with subscript ``in'' are
inflow Jovian parameters. $\mathbf{B}_{G}$ is the Ganymede's dipole field
strength at equitorial surface of the Ganymede. It determines the Ganymede's
dipole field by
$\mathbf{B}_{0}=\left[3\left(\mathbf{r}\cdot\mathbf{B}_{G}\right)\mathbf{r}-\mathbf{B}_{G}r^{2}\right]/r^{5}$.
Other parameters include $\gamma=5/3$ and $\mu_{0}=4\pi\times10^{-7}N/A^{2}$.}

\centering{}%
\begin{tabular}{c|c}
\hline Parameter & Value\\
\hline $\rho_{in}\left[amu/cm^{3}\right]$ & 56\\
$\mathbf{v}_{in}\left[km/s\right]$ & $\left(140,0,0\right)$\\
$p_{in}\left[nPa\right]$ & 3.8\\ $\mathbf{B}_{in}\left[nT\right]$ &
$\left(0,-6,-77\right)$\\ $\mathbf{B}_{G}\left[nT\right]$ &
$\left(-18,51.8,-716.8\right)$\\ $c\left[km/s\right]$ & 6000\\ $m_{i}/m_{e}$ &
25\\ $p_{i}/p_{e}$ & 5\\
\hline
\end{tabular}
\end{table*}

\section{\label{sec:Results}Simulation Results}

\subsection{Overview of basic results}

Before dealing with specific non-MHD physics introduced by the 10-moment
model, it is crucial to make sure that this model correctly captures essential
characteristics of the Ganymede's magnetosphere, as well as key features of
magnetic reconnection.

\subsubsection{Global structure of Alfv\'en wings}

\begin{figure*}
\includegraphics[width=1\textwidth]{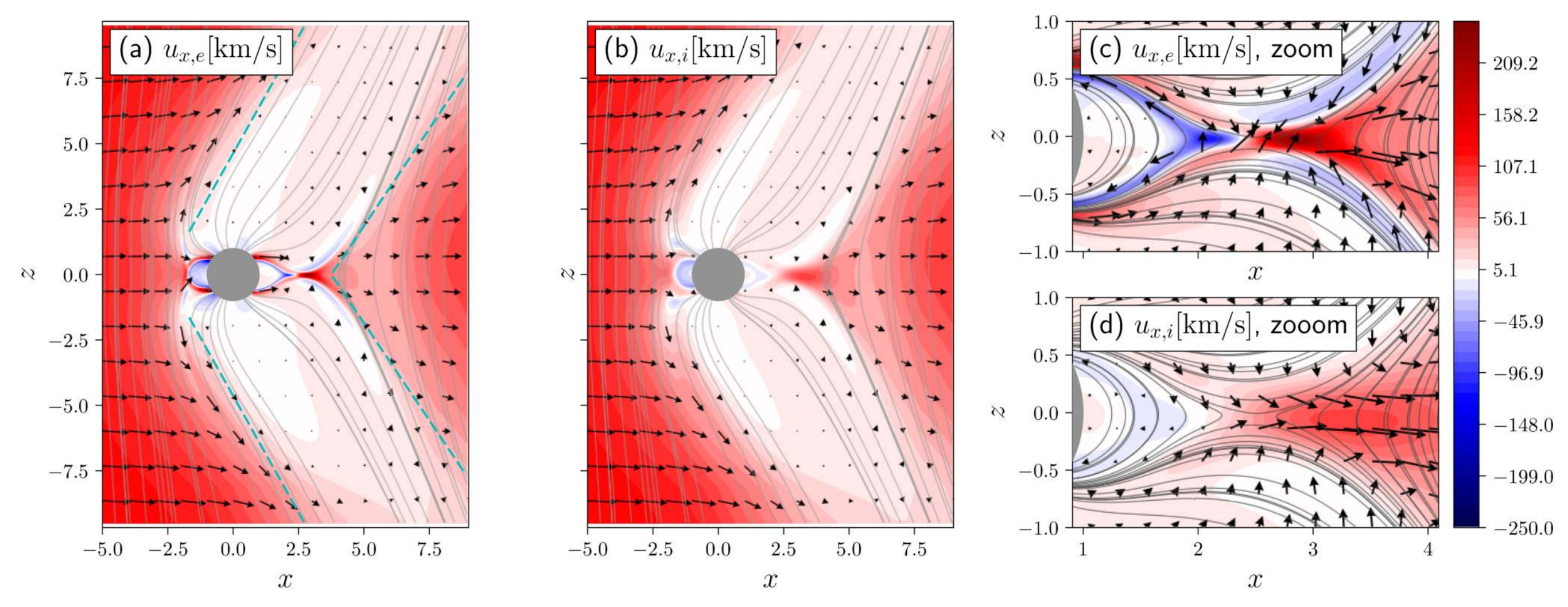}

\caption{\label{fig:xz-plane-vx-by}Meridional cuts of electron (panel a and
c) and ion (panel b and d) flow speeds along $x$ as colored contours.
   $u_{x,e}$ and $u_{y,e}$ can reach $\pm400{\rm km/s}$ and $\pm160{\rm
   km/s}$, respectively, but the color map limits are cut and set to be the
   same for all panels for visualization purpose. Black arrows represent
   in-plane flows for the two species. Black curves are in-plane magnetic
   field line. The cyan dashed lines in panel (a) roughly mark the boundaries
   of the Alfv\'en wings.}
\end{figure*}

A prominent feature of Ganymede is its persisting Alfv\'en wing structure. It
results from the fact that the relative speed of Jovian plasma inflow is
sub-Alfv\'enic and sub-sonic, thus no bow shock is formed ahead of the moon.
Instead, the plasma is slowed down significantly within a pair of tube-like
structures, called the Alfv\'en wings, that extend at an angle to the ambient
magnetic field for a long distance. The Alfv\'en wing structure is clearly
formed in our simulation. Figure \ref{fig:xz-plane-vx-by} shows the meridional
cuts of the electron and ion flow speeds along $x$ \added{direction}. The flow
speeds are dramatically decreased from the Jovian value $140\mathrm{km/s}$ to
as low as $\sim10\mathrm{km/s}$ in the Alfv\'en wings. The dashed lines mark
boundaries of the wings indicated by the gradient in color contours as well as
the diverted flows and abrupt bending of magnetic field lines. The overall
Alfv\'en wing structure is consistent with previous simulation results using
various models \citep{Paty2004,Dorelli2015,Toth2016}. In addition, the figure
also shows the reconnection outflow jets along $\pm x$ at the downstream
X-line near $x\approx2.5R_{G}$ magnified in panels (c) and (d). Ions are
ejected at a bulk speed $\sim160{\rm km/s}$, while electrons are ejected at a
higher speed, about $400{\rm km/s}$. Compared with values identified in the
previous MHD-EPIC study by \citep{Toth2016}, the maximum ion outflow velocity
$u_{x,i}$ is similar, but the maximum electron outflow velocity $u_{x,e}$ is
slower by half, likely due to the larger relative electron mass used in our
study (we used $m_{e}=m_{i}/25$ instead of $m_{i}/100$).

\subsubsection{Reconnection current density carrier}

\begin{figure*}
\includegraphics[width=1\textwidth]{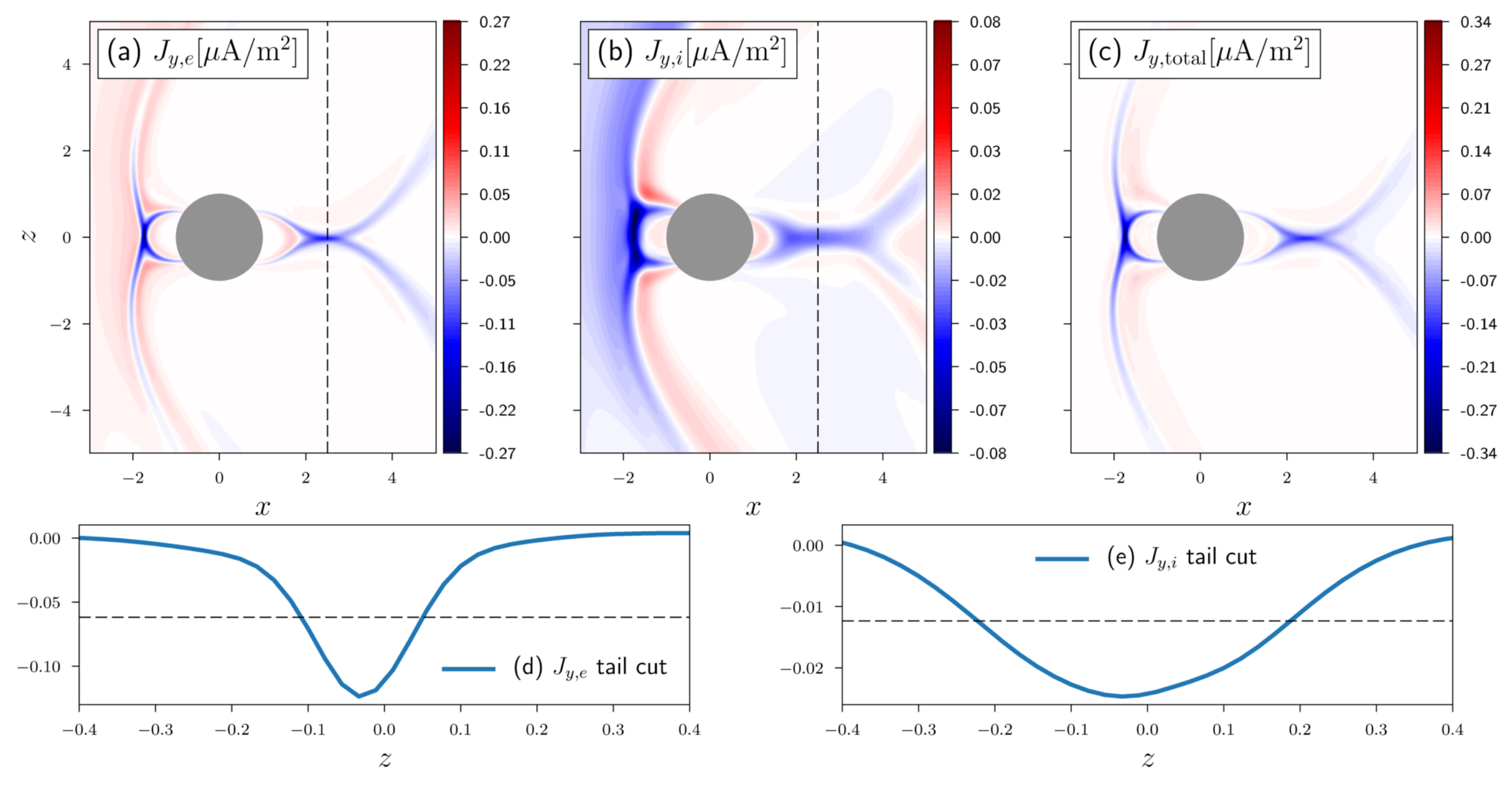}

\caption{\label{fig:Jy_xz-plane}Top, panel (a) to (c): Out-of-plane electron,
ion and total current densities in the meridional plane. Bottom, panel (d) and
(e): $z$-cuts of electron and ion current densities across the tail current
sheets. The cuts are taken at $x=2.5R_{{\rm G}}$ as marked by the dashed lines
in panel (a) and (b). The horizontal dashed lines mark $50\%$ maximums of
$J_{y,e}$ and $J_{y,i}$.}
\end{figure*}

The reconnection physics can be further investigated by looking at the
reconnection-driven current densities. Figure \ref{fig:Jy_xz-plane} gives a
close look at the out-of-plane electron, ion and total current densities in
the meridional plane. Comparing panels (a) and (b), there is a clear
separation of thicknesses for the dense electron and ion current sheets, since
they  scale with electron and ion inertia lengths, respectively. The total
current is carried mainly by the lighter electrons consistent with the
theoretical expectation. The current sheet extends along the field line
separatrices on one end far into the wings, and on the other end down to the
Ganymede's surface. Both $J_{y,e}$ and $J_{y,i}$ are finite upstream of the
magnetopause at about $x=-1.8R_{{\rm G}}$, but they cancel each other in this
region. No plasmoid generation / FTE formation were observed in this
particular simulation. This could be caused by insufficient resolution in
electron kinetic scales below which thin current sheets can break into flux
tubes, and/or that the uniform $k_{e,i}$ chosen might not best fit the local
wave length scales and effectively damped microinstabilities responsible for
FTE formation. The size of the magnetosphere is slightly smaller than obtained
in previous studies, but is still reasonable. The bottom panels (d) and (e)
are cuts of $J_{y,e}$ and $J_{y,i}$ at the downstream side reconnection site.
The cut direction is across the current sheet approximately at $x=2.5R_{{\rm
G}}$ (and $y=0$) as marked by the vertical dashed lines in panel (a) and (b).
The horizontal dashed lines mark the half-maximum locations, which were used
to calculate the FWHM (full width half maximum) current sheet thickness. Thus
the electron current sheet thickness is around $0.15R_{{\rm
G}}\sim3.8d_{e,{\rm Jovian}}$ and the ion current sheet thickness is about
$0.4R_{{\rm G}}\sim2d_{i,{\rm Jovian}}$ , where $d_{e,{\rm Jovian}}$ and
$d_{i,{\rm Jovian}}$ are inertia length based on the upstream Jovian plasma
density. \replaced{This again is consistent with general understanding of
current sheet scales in reconnection.} {This again is consistent with the
general kinetic picture that electron and ion current sheet thicknesses are
mainly characterized by electron and ion scales, respectively (though the
exact scaling laws have not been generally identified) \citep{Wang2015a}.}

\subsubsection{Comparison with in-situ measurements}

\begin{figure*}
\includegraphics[width=1\textwidth]{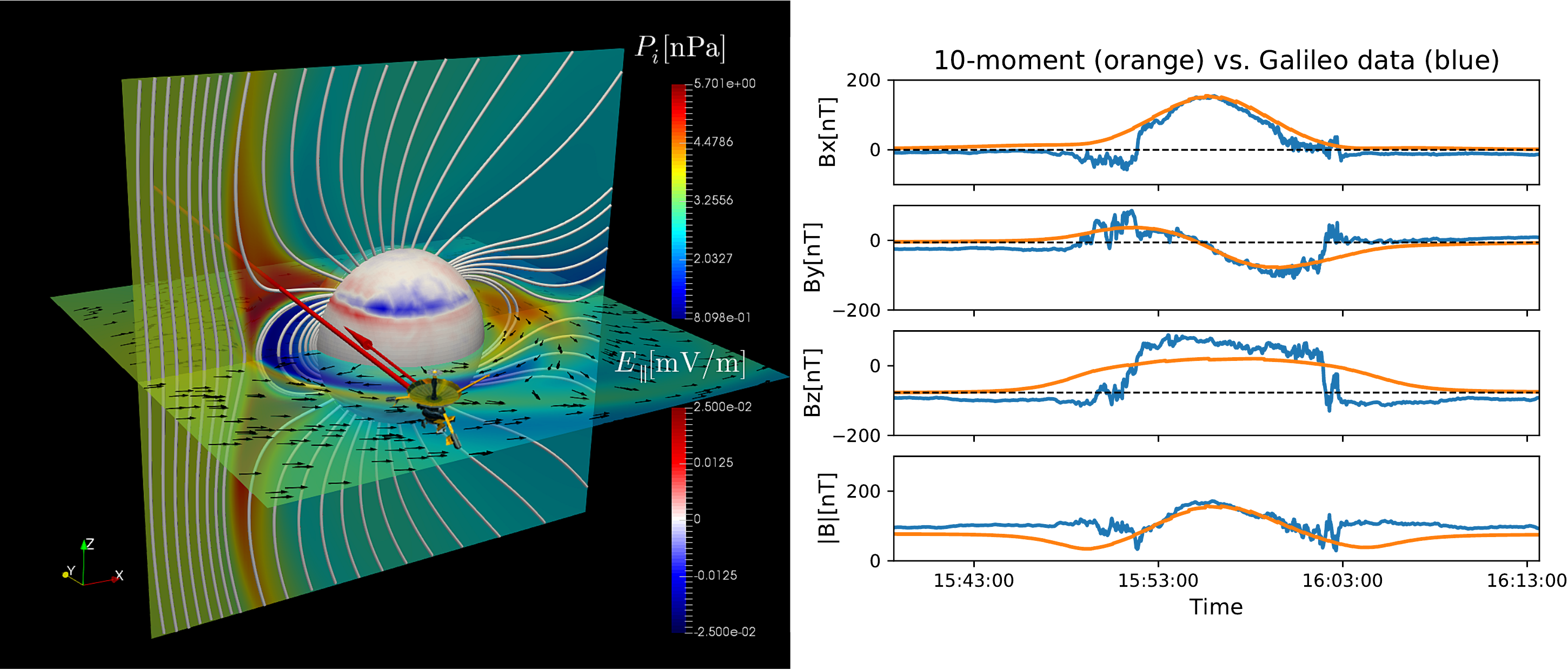}\caption{\label{fig:g8}Left
panel: The red line and arrow indicate the trajectory and direction of the G8
flyby of the Galileo satellite. The cut views of ion pressure $P_{i}$ and
projected magnetic field lines, moon surface electric field $E_{\parallel}$,
and ion plasma flow (arrows) are shown to better demonstrate the location of
the trajectory. Right panel: Comparison of magnetic field data from our
simulation with in-situ measurements by the Galileo magnetometer along the G8
flyby trajectory. The four sub-panels are the three components and magnitude
of the observed (blue curves) and simulated (orange curves) magnetic field.}
\end{figure*}

Figure \ref{fig:g8} shows the simulated and observed magnetic field data along
the published G8 flyby trajectory of the Galileo satellite (depicted by the
red line in the left panel of Figure \ref{fig:g8}). The orange solid lines are
\replaced{stead}{steady}-state modeled values and the blue lines are observed
values. The $x$ and $y$ components of the simulated magnetic field agree well
with the observed within the majority of the magnetosphere. The simulated
$B_{z}$, however, is too flat within the magnetosphere compared to the
spatially enhanced values from observation. In addition, the modeled field
components do not capture the fluctuations at crossings of the magnetosphere
(near 15:50:00 and 16:02:00), either.

We do not \replaced{cliam}{claim} to have achieved perfect
simulation-observation agreement, noting the fact that the field inside the
magnetosphere is dominated by the dipole field, anyway. Previous studies
\citep{Paty2006,Dorelli2015,Toth2016} carried out more sophisticated
improvements and achieved \replaced{more impressive}{better}
agreement. The discrepancies could result from a few factors: 1) The simplified inner boundary
conditions might generate a slightly undersized magnetosphere;
\added{Particularly, the series of papers by Jia et al. showed the importance
of placing a layer below the moon's surface, from $1.05 R_{\rm{G}}$
down to $0.5 R_{\rm{G}}$, in refining the agreement with observations
\citep{Jia2009,Jia2010}. Such a layer is not used in our work (nor in
the Hall MHD work by \citet{Dorelli2015} and the MHD-EPIC work by
\citet{Toth2016}), hence the poorer agreement.} 2) The simulation parameters
do not truly represent the background plasma and field; For example, the
ambient $B_{z}$ from Galileo observation is slightly greater in magnitude than
what we chose ($-77nT$) \added{to be consistent with prior simulation studies}; 3)
\citet{Dorelli2015} and \citet{Toth2016} adjusted the measuring trajectory
slightly;
\deleted{{[\textit{Dorelli et al.}, 2015]} also performed transformation to
local boundary norm coordinates (BNC) regarding the magnetopause normal; These
adjustments are beyond the scope of this paper thus are not performed here.}
4) Previous MHD-EPIC simulation suggested that the fluctuation across the
   magnetopause might be caused by FTEs, which are not observed in our
   simulation for multiple potential reasons as we discussed earlier.
   Nevertheless, consider the relatively simple boundary conditions we
   employed, the agreement achieved here appears to be reasonably good. The
   implementation of more realistic inner boundary conditions, as well as the
   physics of FTEs will be left for future work.

\subsection{Pressure tensor effects in collisionless magnetic reconnection}

In this section, we focus on the role of electron pressure tensor effect in
collisionless reconnection. In a fully kinetic picture, particles are
demagnetized near the X-line where the magnetic field is weak, and undergo
complex meandering motion within the diffusion region
\citet{Vasyliunas1975,Cai1997,Bessho2014a,Bessho2016,Zenitani2016}.
Collectively, these population can drift away from an isotropic and gyrotropic
state to form structured distributions in phase space. In the fluid picture,
this can lead to pressure anisotropy (unequal diagonal elements) and/or
pressure non-gyrotropy (non-vanishing off-diagonal elements) in the pressure
tensor. The 10-moment model being a fluid model, motion of discrete paticles
and hence the full kinetic distribution function is not tracked. However, our
model does allow for the full pressure tensor to evolve according to equation
(\ref{eq:10m-momentum}), incorporating effects of kinetic origin.

\subsubsection{Spatial variation of pressure tensor terms}

\begin{figure*}
\includegraphics[width=1\textwidth]{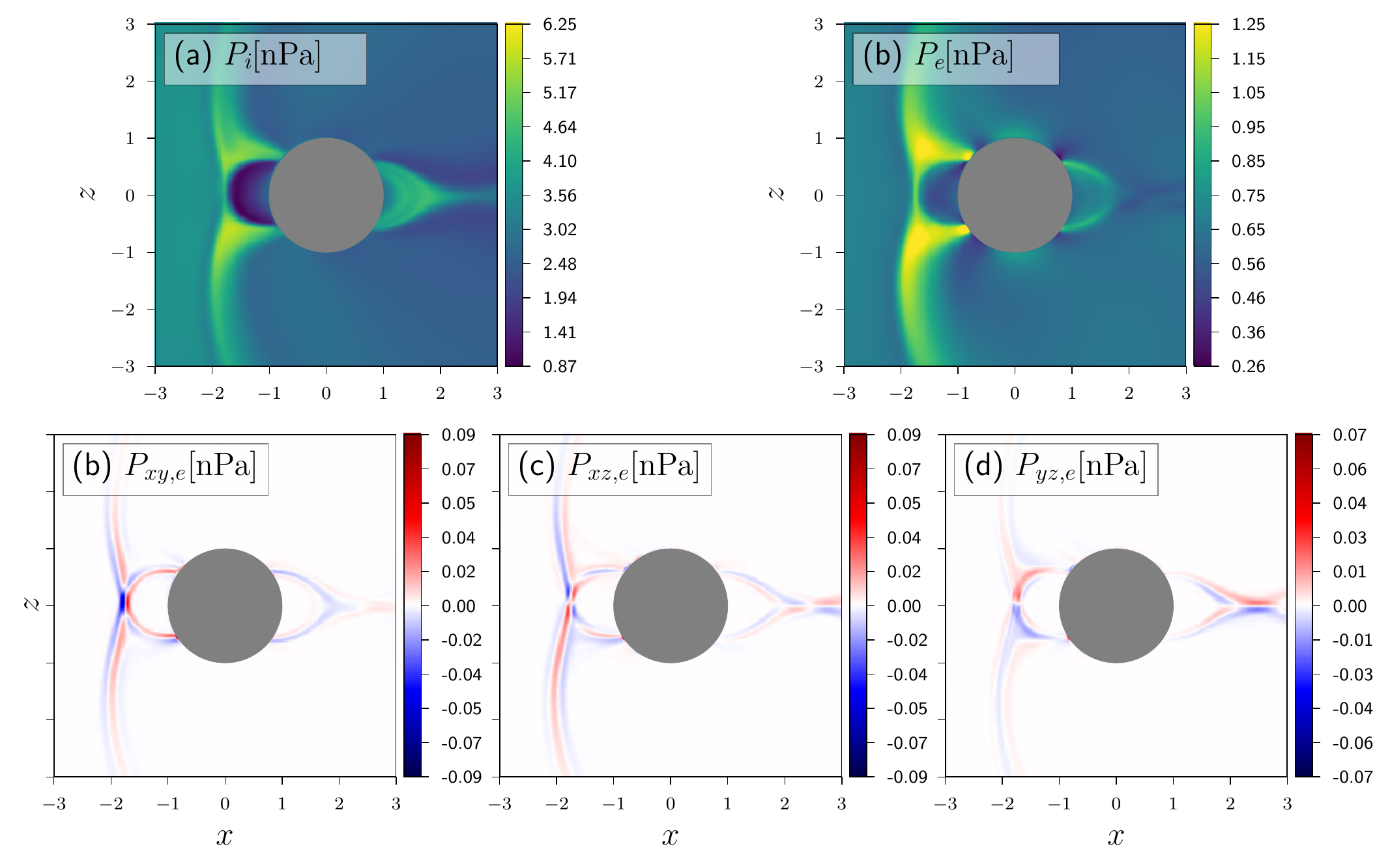}

\caption{\label{fig:pressure-tensors}Meridional cuts of ion and electron
pressure tensor elements at $t=840s$.}
\end{figure*}

Figure \ref{fig:pressure-tensors} shows meridional cuts of the pressure tensor
elements when the simulation has entered an approximately steady state. The
ion scalar pressure, defined by
$P_{i}=\frac{1}{3}\left(P_{xx,i}+P_{yy,i}+P_{zz,i}\right)$, is enhanced near
the separatrices due to the compression of reconnection driven flows. The
spatial pattern of $P_{i}$ is consistent with previous results of
\citep{Toth2016} (Figure 6). The electron scalar pressure $P_{e}$ is enhanced
in more localized regions around the separatrices, and is highest at foot
points of the field lines on Ganymede's surface to form a bright aurora at
latitudes $\sim\pm45^{\circ}$. Non-vanishing off-diagonal elements of the
pressure tensor are created along the separatrices where kinetic microphysics
is important, while remaining zero elsewhere. The polarities of off-diagonal
elements near the X-lines are consistent with previous 2D PIC, Vlasov, and
10-moment simulations \citep{Kuznetsova2001a,Schmitz2006a,Wang2015a} of
anti-parallel reconnection when transformed to the corresponding coordinate
systems. The magnitudes of these elements are of order $\sim0.01{\rm nPa}$,
considerably smaller than those of diagonal elements (of order $\sim1{\rm
nPa}$). However, the sharp spatial gradient associated with the opposite
polarities lead to substantial contribution to the Ohm's law through the
$\nabla\cdot\mathbf{P}_{e}$ term, to be shown later.

\subsubsection{Quantification of non-gyrotropy}

\begin{figure*}
\includegraphics[width=1\textwidth]{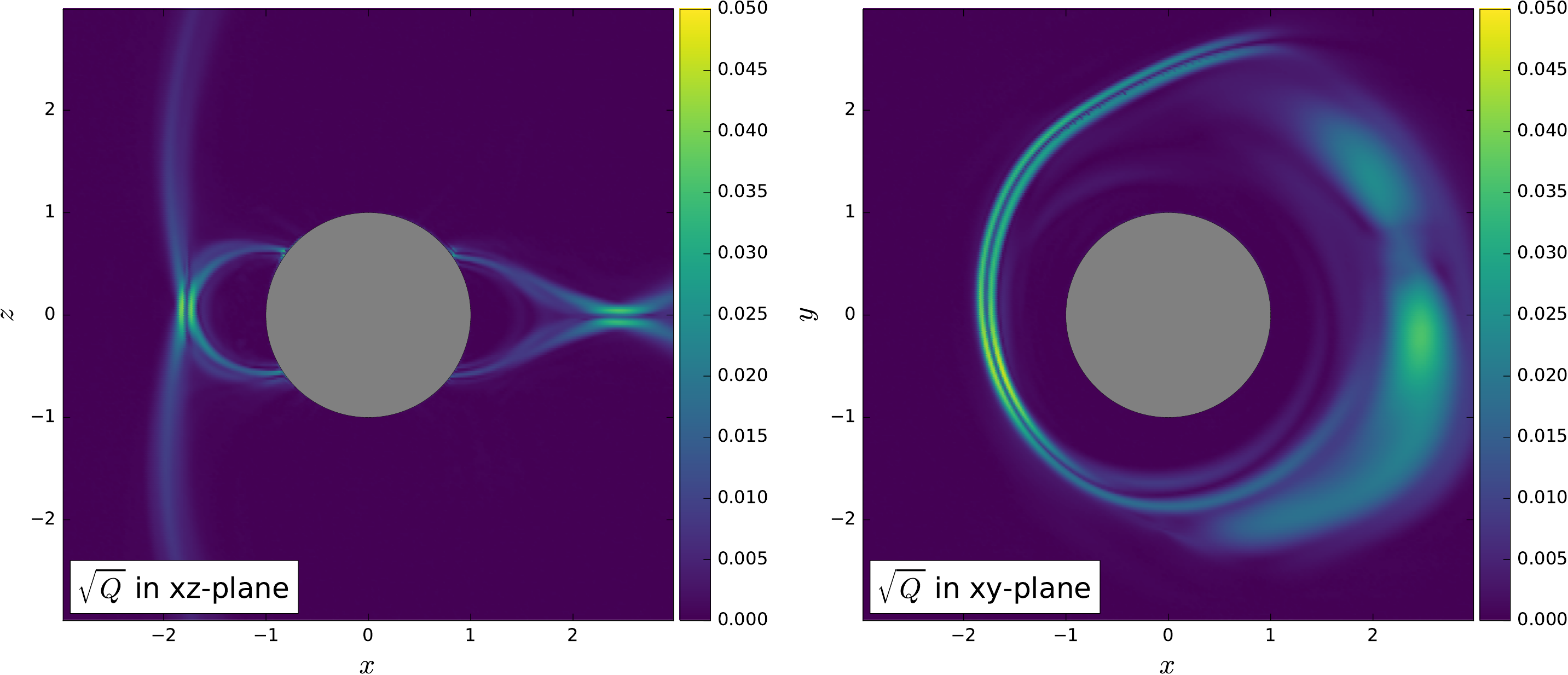}

\caption{\label{fig:sqrtQ}Meridional and equatorial cuts of the non-gyrotropy
measure $\sqrt{Q}$ suggested by \citet{Swisdak2016}. Measured at $t=840s$. $\sqrt{Q}$
can vary between 0 for an entirely gyrotropic state and 1 when off-diagonal
elements are equal to diagonal elements.}
\end{figure*}

A quantitative evaluation of the non-gyrotropy is given in Figure
\ref{fig:sqrtQ} showing the measure $\sqrt{Q}$ suggested by
\citep{Swisdak2016}:
\begin{equation}
Q=1-\frac{4I_{2}}{\left(I_{1}-P_{\parallel}\right)\left(I_{1}+3P_{\parallel}\right)}
\end{equation} where $I_{1}=P_{xx}+P_{yy}+P_{zz}$,
$I_{2}=P_{xx}P_{yy}+P_{xx}P_{zz}+P_{yy}P_{zz}-P_{xy}^{2}-P_{xz}^{2}-P_{yz}^{2}$,
and $P_{\parallel}=\hat{\mathbf{b}}\cdot\mathbf{P}\cdot\hat{\mathbf{b}}$.
Here, we have neglected subscripts $e$ that represent electrons. As shown in
the meridional view, $\sqrt{Q}$ is intensified near the X-line and along the
two ends of open-closed field line separatrices, with maximums just upstream
of the X-lines. $\sqrt{Q}$ drops at center of the X-line since the
off-diagonal elements drop to zero due to asymmetry. At the time shown, the
maxima at the upstream and downstream\textcolor{red}{{} }side X-line are about
$0.05$ and $0.03$ respectively, smaller than but still comparable to those
found in fully kinetic simulations \citep{Swisdak2016,Zenitani2016}. The
greater $\sqrt{Q}$ on the upstream side might result from its asymmetric
reconnection configuration, consistent with the conclusion in
\citep{Swisdak2016} that asymmetric reconnection produced a larger $\sqrt{Q}$
than a comparable symmetric reconnection setup. The substantial non-gyrotropy
$\sqrt{Q}$ formed in the simulation indicates that the non-gyrotropic pressure
effect cannot be neglected.

\subsubsection{Decomposition of the Ohm's law}

\begin{figure*}
\includegraphics[width=1\textwidth]{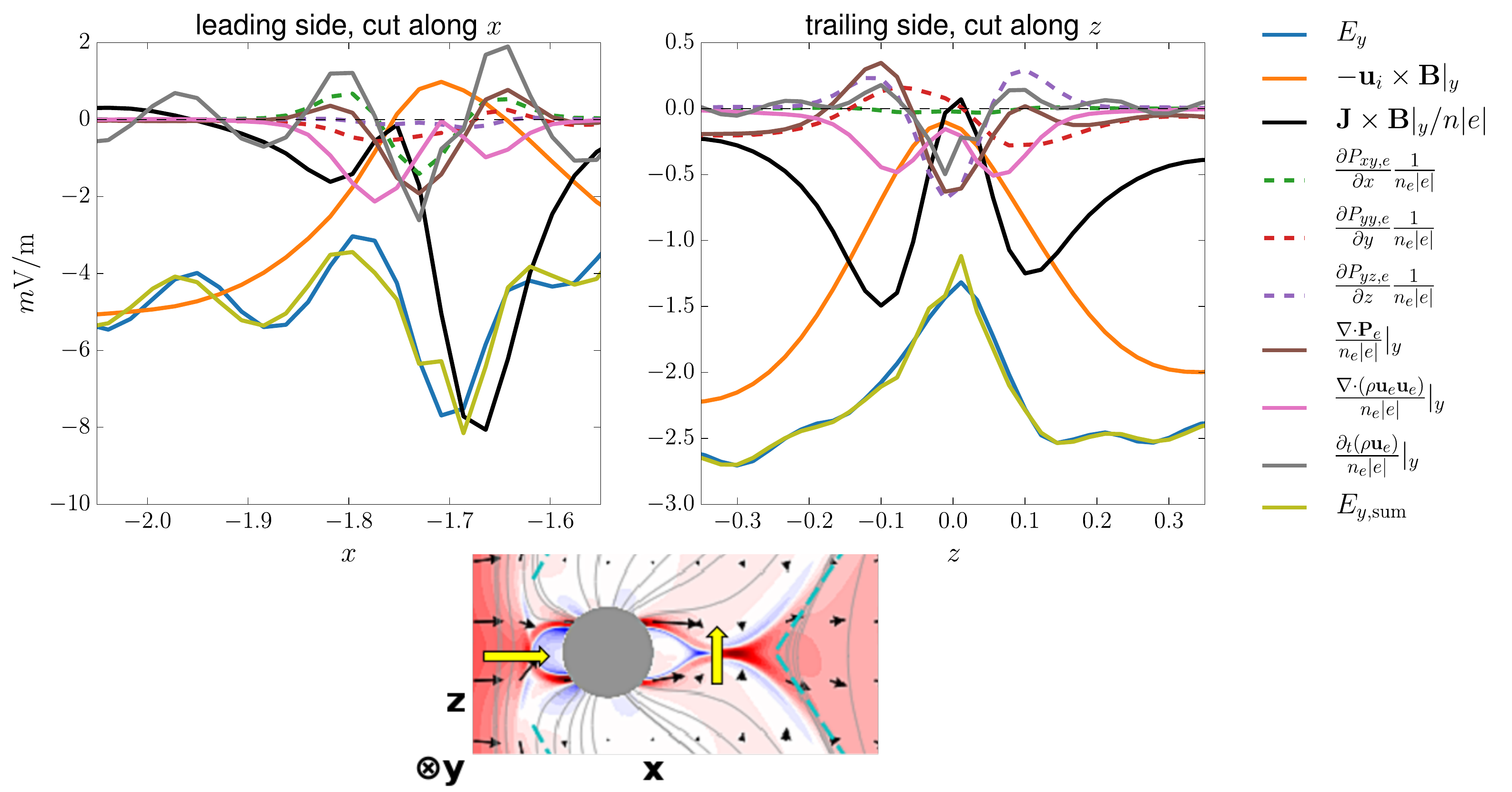}

\caption{\label{fig:ohm}Top row: 1D cuts of terms in the generalized Ohm's law
across upstream (left) and downstreem (right) side current sheets.  Only the
components out of the reconnection plane, that is, $y$-components are shown.
The elements in the $\nabla\cdot\mathbf{P}_{e}$ term are represented by dashed
lines. The thick, yellow arrows in the bottom panel indicate directions of the
cuts which are along the inflow directions, i.e., along $x$ direction for the
upstream side, and along $z$ direction for the downstream side.}
\end{figure*}

Finally, we examine the contribution of $\nabla\cdot\mathbf{P}_{e}$ in the
generalized Ohm's law. Figure \ref{fig:ohm} shows 1D line cuts of terms in the
Ohm's law along the inflow direction at the upstream and downstream side
reconnection sites. We only show terms in the following $y$ component of the
generalized Ohm's law,

\begin{eqnarray} & E_{y}= &
 (-u_{z,i}B_{x}+u_{x,i}B_{z}) +\frac{1}{n_
 {e}\left|e\right|}(J_{z}B_{x}-J_{x}B_{z}) \nonumber\\&&-\frac{1}{n_
 {e}\left|e\right|}\left(\frac{\partial P_{xy,e}}{\partial x}+\frac{\partial
 P_{yy,e}}{\partial y}+\frac{\partial P_{yz,e}}{\partial z}\right)\nonumber \\
 &&
 -\frac{1}{n_{e}\left|e\right|}\left[\frac{\partial\left(\rho_{e}u_{y,e}\right)}{\partial
 t}+\nabla\cdot\left(\rho\mathbf{u}_{e}u_{y,e}\right)\right].\label{eq:ohm-electron-y}
\end{eqnarray} The directly measured $E_{y}$ (blue curve) agrees well with
$E_{y,sum}$ (olive curve), the summation of terms on the right hand side of
equation (\ref{eq:ohm-electron-y}). For simplicity, the cuts are made across
the stagnation point. Overall, the electric field and the decomposed
components are highly structured in space. \replaced{Away from the stagnation
point, i.e., from a distance further than electron inertia scale, $E_{y}$ is
supported mainly by the convection component
$-\mathbf{u}_{e}\times\mathbf{B}$. Near the stagnation point, the convection
component drops to near zero and other sources become important.}{Away from
the stagnation point, i.e., from a distance further than the ion inertia length,
$E_{y}$ is supported mainly by the convection component
$-\mathbf{u}_{i}\times\mathbf{B}$. For the fairly symmetric downstream
reconnection, the Hall term (the black curve) is important in the
``shoulder'' regions on the two sides of the $x$-line where ions are
demagnetized but electrons remain largely magnetized. This is consistent with
previous understanding of symmetric reconnection (see, e.g.,
\citep{Wang2015a}). For the upstream crossing, the asymmetry brings in
substantially more complexity. For example, the Hall term is smaller on the
left side shoulder than on the right side shoulder. Also, the convection term
$-\mathbf{u}_i\times\mathbf{B}$ passes zero two times, whereas the left
passing near $x=-1.75R_G$ is near the null point and the right passing
is near
$x=-1.625R_G$ is near the stagnation point. Similar asymmetries occur in local
simulations of asymmetric reconnection, too, in qualitatively consistent ways
(see, e.g., \citep{Cassak2007b}).

Right at the reconnection sites, both the convection term and the Hall term
vanish since flow velocities and magnetic field both approach zero, while the
$-\nabla\cdot\mathbf{P}_{e}/n_{e}\left|e\right|$ term starts to play an
important role in supporting $E_y$.} For the highly asymmetric upstream side
reconnection, the $\partial P_{xy,e}/\partial x$ term is significant, the
$\partial P_{yy,e}/\partial y$ that results from the diagonal element gradient
is substantially smaller, and the $\partial P_{yz,e}/\partial z$ term is
almost negligible. For the fairly symmetric downstream side reconnection, the
$\partial P_{xy,e}/\partial x$ term is negligible, and the $\partial
P_{yy,e}/\partial y$ and the $\partial P_{yz,e}/\partial z$ terms are
significant and comparable in magnitude. The difference in contributing
off-diagonal element is due to fact the components are not measured in the
same consistent coordinate system regarding the ``ambient'' magnetic field,
which is along $\pm z$ for the upstream side and $\pm x$ for the downstream
side. If we rotate the coordinate by $90^{\circ}$ in the $xz$ plane for the
upstream side, than we would find consistent results. Nevertheless, it is
clear that not only the off-diagonal elements, but also the diagonal elements
can contribute to the Ohm's law in 3D. Particularly, at the downstream side,
the scalar gradient term $\partial P_{yy,e}/\partial y$ contribute
substantially to $E_{y}$ away from the stagnation point at the time shown. On
the other hand, the electron inertia is also important as displayed by the
substantial contribution from the flow divergence term
$\nabla\cdot\left(\rho_{e}\mathbf{u}_{e}u_{y,e}\right)$ and the time
derivative term $\partial\left(\rho\mathbf{u}_{e}\right)/\partial t$.

\subsection{Asymmetric patterns of electrons and ions}

Previous MHD-based simulations with single or multiple ion species showed
highly asymmetric patterns in surface brightness and bulk flows around the
Ganymede \citep{Paty2004,Paty2006,Dorelli2015}. However, electron dynamics
were derived from assumptions instead of directly evolved in those models
since they assume $m_{e}=0$. In the 10-moment model, electrons are
incorporated as a separate, independent fluid. We will see that electrons also
demonstrate strong spatial variability and contribute significantly to the
overall asymmetry due to its negative charge and small, but finite inertia.

\subsubsection{Electron vs. ion drift belts}

\begin{figure*}
\includegraphics[width=1\textwidth]{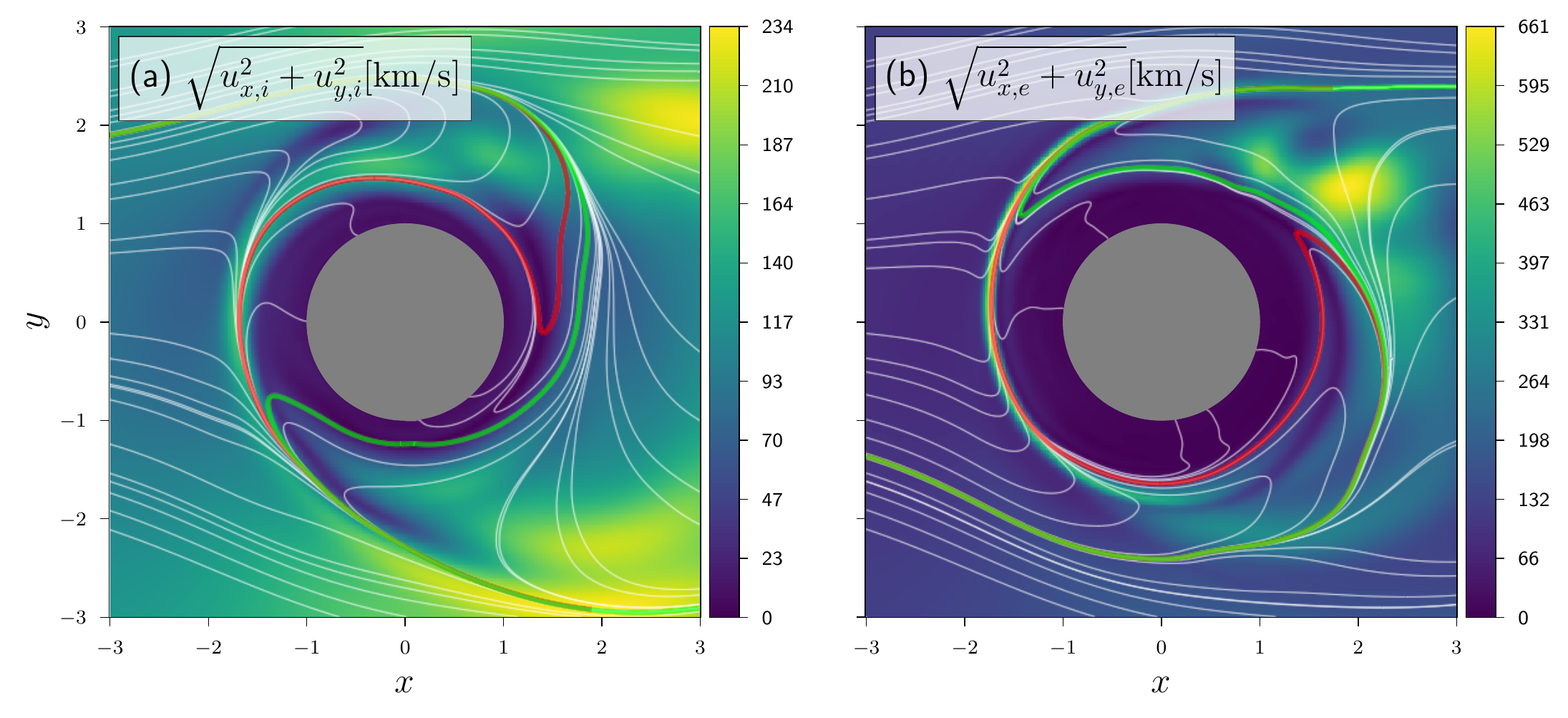}

\caption{\label{fig:2d-drift}Equatorial cuts of electron (left) and ion (ion)
streamlines color coded by the in-plane speed of each species.}
\end{figure*}

We first look at the equatorial cuts in Figure \ref{fig:2d-drift} of electron
and ion streamlines over color contours for in-plane speeds. The reconnection
electric field and diamagnetic drift at the reconnection sites drives the two
species in opposite directions, that is, clockwise for the electrons and
anti-clockwise for the ions when viewed from $+z$ direction. The drifts then
interact with the incident Jovian inflows to produce very different drift
patterns for the two oppositely charged species. Consequently, the locations
where the Jovian electron and ion flows enter the ``inner magnetosphere'' are
different. For example, the thickened red lines mark the streamlines that can
marginally go around the moon and exit or enter the inner magnetosphere. Also,
for both species, the flow patterns are highly structured and sheared. As a
result, the streamlines chosen and marked in red and green can be diverted to
follow very different paths though they originated from almost the same
locations in the Jovian plasma. It also worth noting that, for both species,
the flows are accelerated along the X-line to fairly high speeds eventually
when they ``escape'' back to the Jovian plasma in the wake at the
Jovian-facing side ($y>0$) for electrons and the other side ($y<0$) for ions.
Finally, at the sub-Jovian flank the returning and upstream flows create
shears, though we do not observe clear signatures of Kelvin-Helmholtz
instability (KHI). Readers are also encouraged to study the 3D image in the
Supporting Information (Figure S1) for perspective views of the complex
electron and ion flow patterns.

\subsubsection{Surface brightness}

High-speed reconnection outflow jets can propagate along the field line
separatrices to hit the Ganymede's surface and form a bright aurora oval. This
process takes place at both upstream and downstream sides of the Ganymede,
during which magnetic field-aligned electric field can efficiently energize
the plasmas to be observable as enhanced brightness of auroral emissions.
Payan et al. developed a sophisticated auroral brightness model in their
multi-ion MHD simulation to account for atomic oxygen emission
\citep{Payan2015a}. Though such brightness model has not been implemented in
our 10-moment multi-fluid model, we can get a qualitative picture by using
plasma pressures directly as a proxy for brightness.

\begin{figure*}
\begin{centering}
\includegraphics[width=1\textwidth]{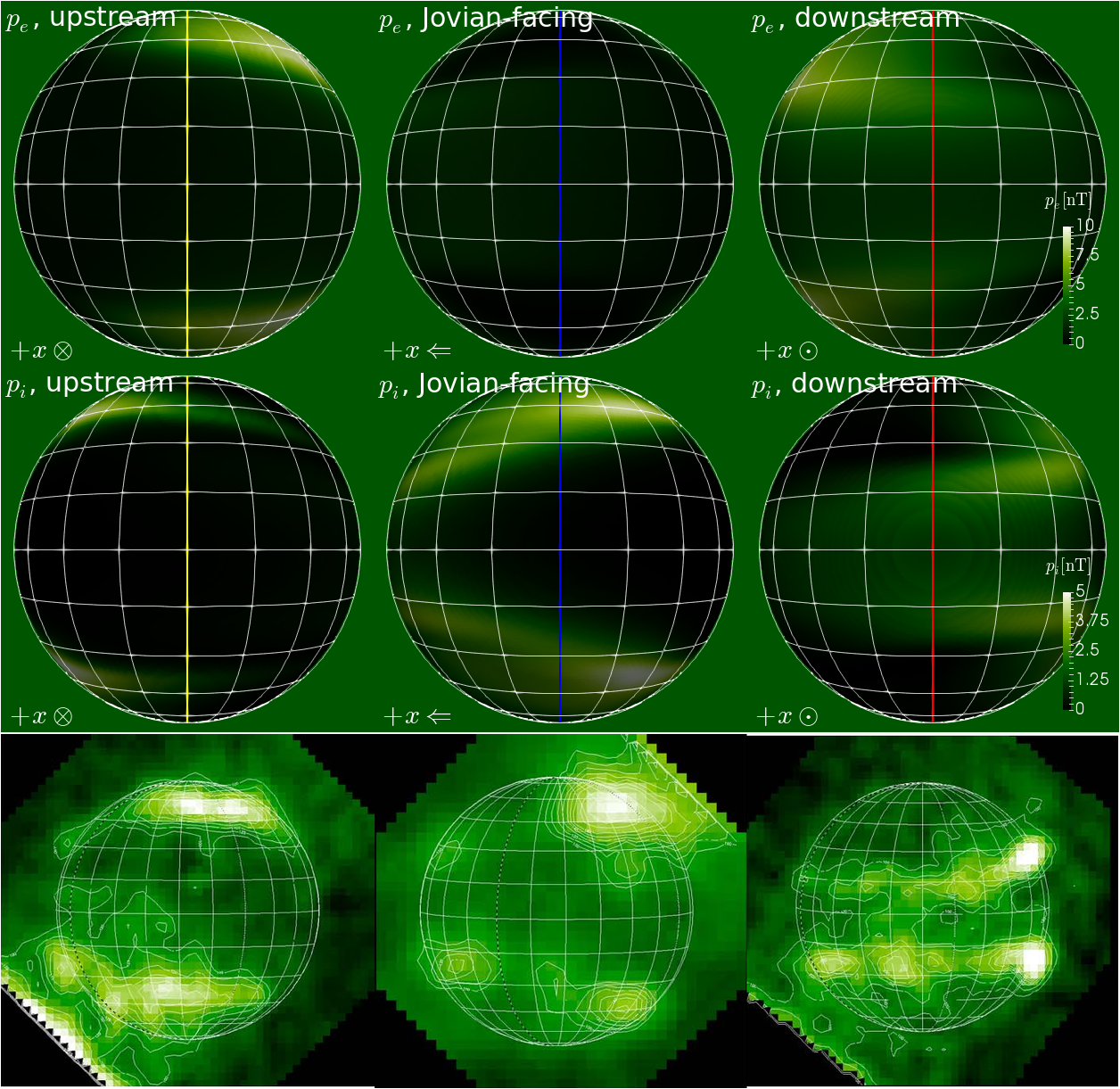}
\par\end{centering}
\caption{\label{fig:surface-pressure}Top and middle rows: Electron (upper row)
and ion (lower row) scalar pressures (in units of ${\rm nT}$) on a sphere
$263\mathrm{km}$ above the Ganymede surface. The three columns are for the
upstream, Jovian-facing and downstream hemispheres. $+z$ points upward in all
panels. $+x$ are into the plane, right to left, and out of the plane in the
three columns, and are marked at left lower corner of each panel. The red,
blue, and yellow curves mark west longitudes $270^{\circ}{\rm W}$,
$90^{\circ}{\rm W}$, $0^{\circ}/360^{\circ}{\rm W}$. Third row: Surface
brightness from oxygen emission observed by HST. The three images are
reproduced from the (c), (b), (a) panels of Figure 2 (upper row) of
\citep{McGrath2013}.}
\end{figure*}

Figure \ref{fig:surface-pressure} shows surface morphologies of electron and
ion scalar pressure ``brightnesses'' on the upstream, Jovian-facing, and
downstream hemispheres, along with the brightness morphologies due to atomic
oxygen emission observed by HST \citep{McGrath2013}. The electron and ion
``brightnesses'' become fairly stable once the simulation has reached a
quasi-steady state, which is consistent with previous observations of
repeatable and relatively stable emission patterns \citep{McGrath2013}. The
pressures on the upstream hemisphere are enhanced at higher latitudes than on
the downstream hemisphere. This is in support of previous hypothesis of
acceleration by reconnection electric field since the upstream cusp region is
at higher latitude than the downstream cusp. However, the enhancement patters
of the two species are very different. For example, on the downstream
hemisphere, $p_{e}$ is mostly enhanced near $165^{\circ}{\rm W}$,
$30^{\circ}{\rm N/S}$, while $p_{i}$ is mostly enhanced near $30^{\circ}{\rm
W}$, $25^{\circ}{\rm N/S}$ (here, following the definition in
\citet{McGrath2013}, the west/east hemisphere corresponds to the dusk/dawn
hemisphere with west longitude $0^{\circ}{\rm W}$ at center the upstream
hemisphere, marked by the yellow curves in Figure \ref{fig:surface-pressure}).
In comparison, the brightest spots observed by HST are located $90^{\circ}{\rm
W}$ on the northern hemisphere, consistent wit the pattern of $p_{i}$ in our
simulation, but $90^{\circ}{\rm W}$ on the southern hemisphere, consistent
with patterns of $p_{e}$. For the Jovian-facing hemisphere, the enhancement in
$p_{e}$ can be barely seen, while $p_{i}$ is clearly enhanced near
$330^{\circ}{\rm W}$, $45^{\circ}{\rm N/S}$, very close to the observation.
For the upstream hemisphere, the observed high emission regions seem to be
consistent with the high $p_{e}$ region in our simulation, but is more
stretched. Over all, our simulation captured some key features, e.g.,
``emissions'' at different latitudes on the upstream and downstream
hemispheres. The real observed patterns seem to be a complication of modeled
electron or ion ``brightnesses''. There remain some discrepancies, which are
probably due to lack of more realistic inner boundary conditions and that we
are modeling only one ``average'' ion species instead of multiple ones.

\begin{figure*}
\includegraphics[width=1\textwidth]{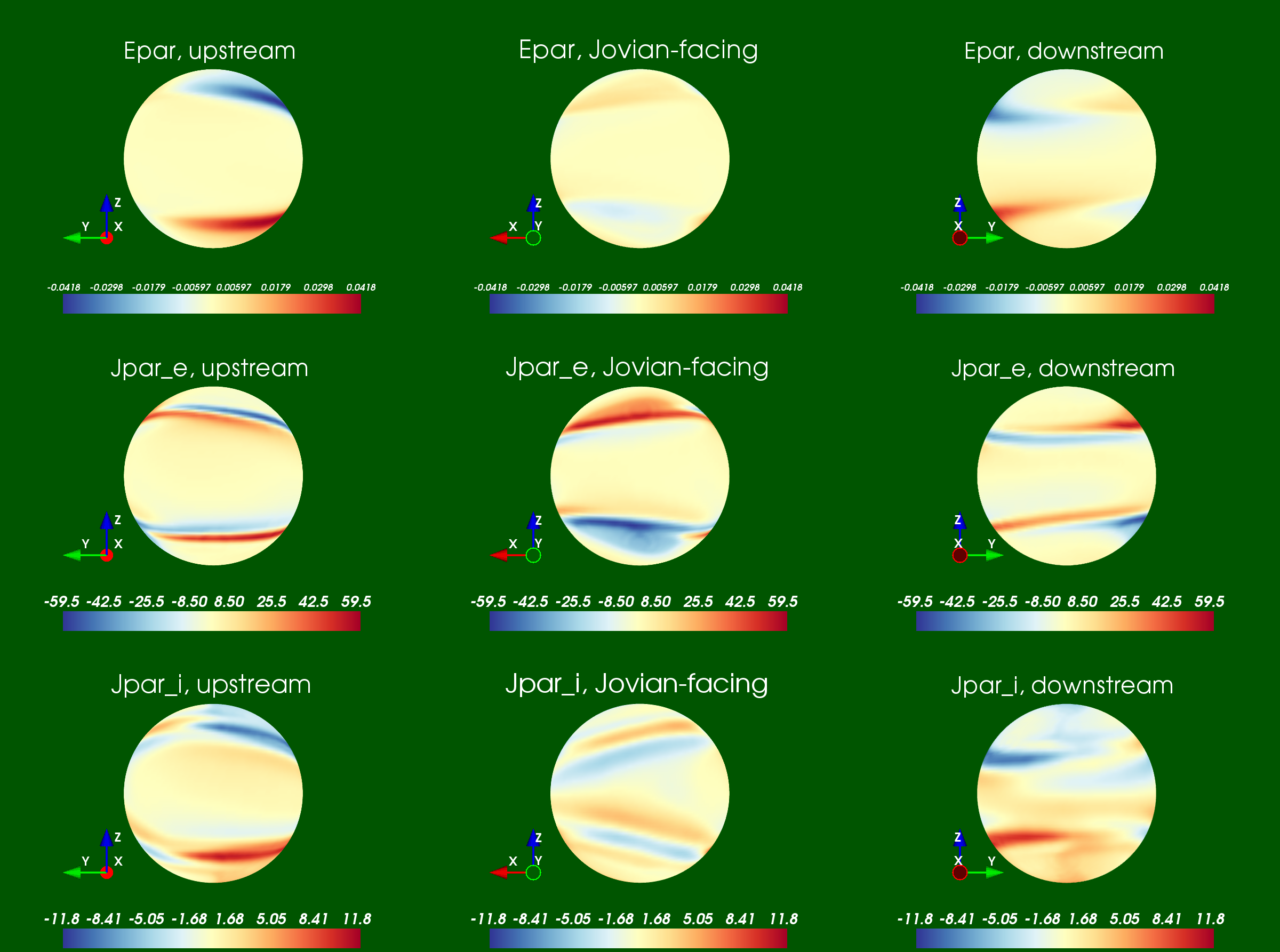}

\caption{\label{fig:surface-Epar}Parallel electric field (top row), electron
current (middle row), and ion current (bottom row) on a sphere
$263\mathrm{km}$ above the Ganymede surface. The three columns are for the
upstream, Jovian-facing and downstream hemispheres, same as
Figure \ref{fig:surface-pressure}.}
\end{figure*}

Next, we look at the possible role of magnetic-field aligned electric field
$E_{\parallel}$ by comparing the electron and ion pressure ``brightness''
surface maps in Figure \ref{fig:surface-pressure} with the corresponding maps
of $E_{\parallel}$ in Figure \ref{fig:surface-Epar} (the top row). It is
clearly shown that $p_{i}$ tends to \added{be} enhanced where $E_{\parallel}$
is positive, while high $p_{e}$ regions tend to correlate with negative
$E_{\parallel}$regions. Such tendency is less clear in the southern hemisphere
map in the middle column (Jovian-facing) that the negative $E_{\parallel}$
band at latitude $-30^{\circ}$ does not seem to correspond to clear
enhancement in $p_{e}$. Nevertheless, the correlation is
\replaced{observed}{observable} overall. It is worth noting that the
$E_{\parallel}$ in our 10-moment simulation is supported by the
self-consistently generated pressure tensors divergence and partially by
electron inertia effects, instead of numerical or artificially prescribed
resistivity. \added{To understand this better, it is also interesting to examine the
role of field aligned currents $J_{\parallel,e}$ and $J_{\parallel,i}$ (middle
and bottom rows of Figure \ref{fig:surface-Epar}) due to the acceleration by
$E_{\parallel}$. The polarities of $J_{\parallel,e}$ and
$J_{\parallel,i}$ are closely related to the pressure enhancement patterns. In
the northern hemisphere, for example, the magnetic field lines are downward
into the moon's surface, thus positive $J_{\parallel,i}$ corresponds to
downward ion flows and thus the enhancement of $P_i$ due to pileup of ions at
the foot-points. Similarly, negative $J_{\parallel,e}$ corresponds to downward
electron flows and hence enhanced $P_e$. In the southern hemisphere, the field
lines are upward thus the logic applies in the opposite way, i.e., positive
$J_{\parallel,e}$ corresponds to $P_e$ enhancement.}

\begin{figure*}
\begin{centering}
\includegraphics[width=1\textwidth]{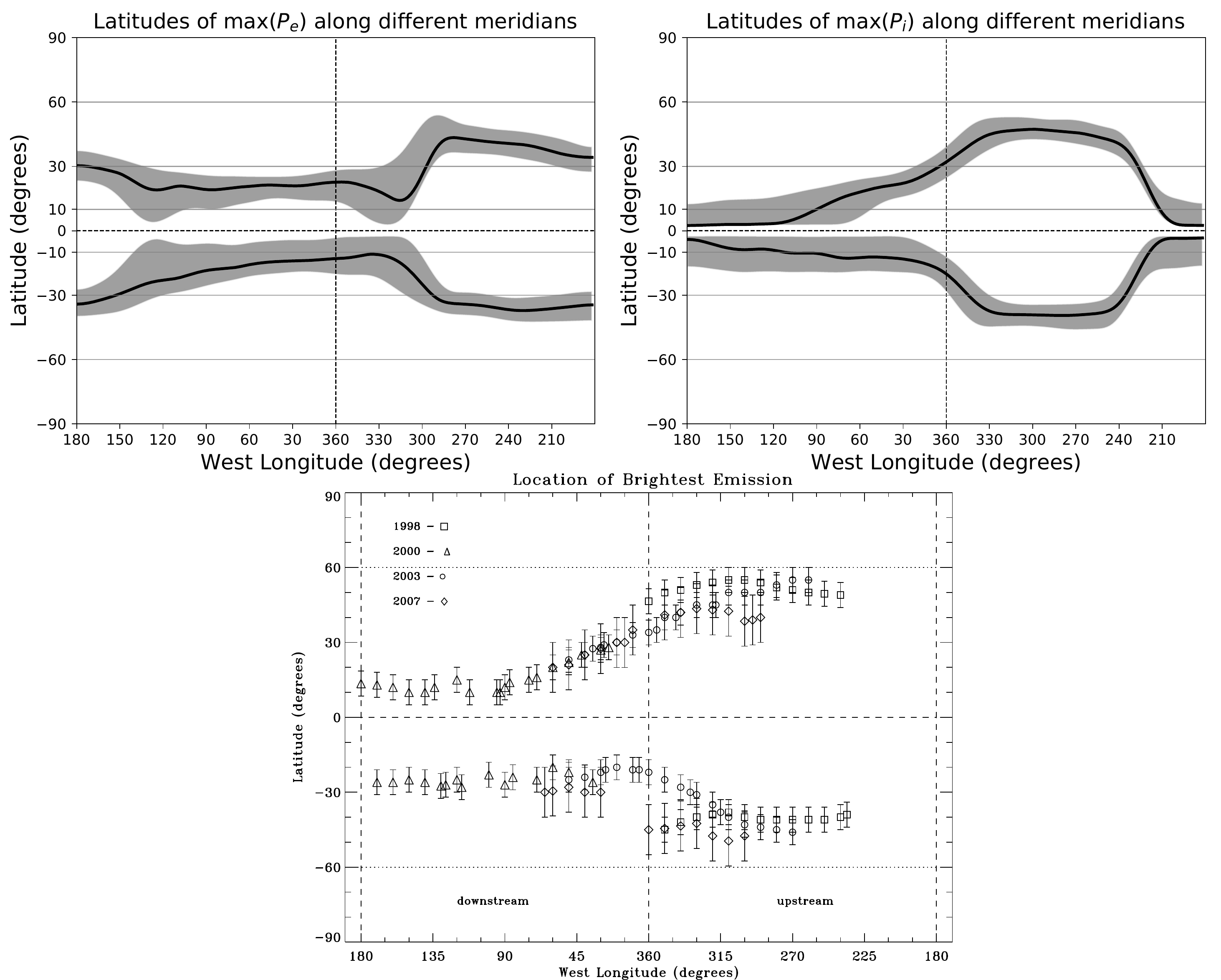}
\par\end{centering}
\caption{\label{fig:brightness-map}Top panels: Locations of peak pressures,
$P_{e}$ and $P_{i}$, on a sphere $263\mathrm{km}$ above the Ganymede's
surface. The solid black curves are the exact locations of maximum values, and
the shaded stripes represent regions of values higher than $90\%$ of the
maximums. Bottom panel: Locations of peak auroral emission brightness observed
by HST in 1998, 2000, 2003, and 2007 (reproduced from Figure 3 of
\citep{McGrath2013}).}
\end{figure*}

Finally, Figure \ref{fig:brightness-map} shows the latitude-longitude maps of
peak brightnesses derived from Figure \ref{fig:surface-pressure}. The peak
electron pressure band in the top panel reaches highest latitude
$\sim\pm50^{\circ}$ abruptly near about longitude $300^{\circ}$ and then drops
fast. The ion band in the middle panel rises much more gradually and stays at
highest latitude over a wider plateau. Both bands, particularly the ion band,
approximately agree with in-situ measurements by HST, reproduced in the bottom
panel of Figure \ref{fig:brightness-map}. They are also consistent with the
Figure 4 of \citep{Payan2015a} from multi-ion MHD simulations with a more
realistic brightness model, particularly the case where the moon is near the
center of the Jovian plasma sheet, which is consistent with our simulation
setup (flyby G8).

\section{\label{sec:Conclusions}Discussion and Conclusions}

We have successfully modeled Ganymede's magnetosphere using the 10-moment
model. This model provides an efficient way to address deficiencies in modern
MHD-based magnetospheric codes, particularly the lack of electron and ion
pressure tensor effects. The model correctly captured key features of the
Ganymede's magnetosphere like the Alfv\'en wing structure and produced
magnetic field data that agree reasonably well with in-situ observation. More
importantly, the simulation clearly demonstrated, for the first time in the
context of realistic 3D magnetosphere, the importance of full electron
pressure tensor in collisionless reconnection. In fact, both the diagonal and
off-diagonal elements are shown to contribute significantly to the Ohm's law
term $\nabla\cdot\mathbf{P}_{e}$ in 3D. The electron inertia terms were also
shown to contribute to the reconnection electric field.

Since the 10-moment model tracks the electron species independently, the
effects introduced by electron can be examined straightforwardly. From our
simulation, electrons are shown to form asymmetric drift patterns very
different from ions, as well as different brightness patterns on the surface
of Ganymede. The brightness maps captures some key features observed by the
HST/ACS-SBC \citep{McGrath2013}, and could be further refined by using more
realistic inner boundary condition and/or more sophisticated brightness model.
While the MHD-EPIC model also evolves electron independently and is capable of
carrying out similar study in principle, the PIC region employed in
\citep{Toth2016} did not cover the vicinity of the Ganymede's surface and such
study was not performed.

Compared to the MHD-EPIC model which also aims at incorporating kinetic
effects in global modeling, the 10-moment model is computationally less
expensive for the same domain size and resolution. The trade-off is that the
10-moment model does not fully retains kinetic effects but truncates at the
2nd order moment. The fluid-representation of the 10-moment model also allows
more straightforward coupling with other fluid-based codes like MHD, while the
MHD-EPIC approach is generally more difficult to implement due to the vast
discrepancy between the MHD model and the PIC model.

One limitation of the 10-moment model is that the inner boundary conditions
are more \replaced{difficulty}{difficult} to set compared to MHD due to the
larger number of state quantities. For example, to obtain stable solutions to
the Ganymede's magnetosphere, we set the electric field $\mathbf{E}$ in inner
boundary ghost cells to zero, effectively allowing $\mathbf{E}$ to
``semi-float''. For a MHD code, boundary condition for $\mathbf{E}$ is not
necessary since it is computed from a prescribed form of the Ohm's law.

This work can be further extended to study multiple topics. For example, our
simulation used only one ``average'' ion species with an average mass and
charge. However, in-situ measurements indicated presence of multiple ion
species with different masses and temperatures. Multi-fluid MHD simulations by
Paty et al. already showed effects of multiple ion species in terms of
convection etc, but did not study how the reconnection rate scales with these
effects \citep{Paty2004,Paty2006,Paty2008}. We plan to incorporate more ion
species and study how the global convection and local reconnection dynamics
scale with the masses and temperatures of various ion sources. On the other
hand, the electron dynamics could be combined with the brightness model
suggested by \citet{Payan2015a} to give potentially better prediction of
aurora emission. Finally, the model may also be applied to other planetary
systems, including Mercury and Earth. For example, the 10-moment model can
couple with the ionospheric module in the Earth code OpenGGCM to give a more
complete picture of magnetospheric physics and possibly more accurate
prediction of space weather events.

\acknowledgments

L. Wang, K. Germaschewski, and J. Raeder are supported by the NASA Grant No.
   NNX13AK31G.
L. Wang and K. Germaschewski are also supported by the DOE Grant No.
   DESC0006670.
A. Hakim and A. Bhattacharjee are supported by the NSF Grant No. AGS-1338944.
C. F. Dong is supported by the NASA Living With a Star Jack Eddy Postdoctoral
   Fellowship Program, administered by the University Corporation for
   Atmospheric Research. Computations were performed on Trillian, a Cray
   XE6m-200 supercomputer at UNH supported by the NSF MRI program under Grant
   No. PHY-1229408. We thank J. C. Dorelli for providing the magnetic field
   data from Galileo observation shown in Figure \ref{fig:g8}, and A. Glocer
   for useful discussion on implementation of inner boundary conditions.  The
   simulation configuration file and output data can be obtained from
   the author L. Wang.


\listofchanges

\end{document}